\documentclass[11pt]{article}

\usepackage[T1]{fontenc}
\usepackage[utf8]{inputenc}
\usepackage{lmodern}
\usepackage[a4paper,margin=1in]{geometry}
\usepackage{amsmath,amssymb}
\usepackage{booktabs}
\usepackage{enumitem}
\usepackage{graphicx}
\usepackage{xcolor}
\usepackage{hyperref}
\usepackage{microtype}

\hypersetup{
  colorlinks=true,
  linkcolor=blue!55!black,
  citecolor=blue!55!black,
  urlcolor=blue!55!black
}

\setlist[itemize]{leftmargin=1.4em}
\title{\vspace{-1.5em}
Attractive and Repulsive\\
Pattern Control in Sequence Generation}
\author{Fran\c{c}ois Pachet}
\date{June 2026}

\begin{document}
\maketitle

\begin{abstract}
Variable-order Markov models are useful for symbolic sequence generation
because they preserve local syntax or style while adapting context length.
Over long continuations, they may enter ``tunnels'':
attractor-like corridors of recurring high-order contexts, repeated suffixes,
or locally periodic passages. This tunneling is already present in
unconstrained Markov continuation, before any belief propagation (BP) or
regular constraints are introduced. Previous BP-Regular work showed that
regular constraints can be compiled into automata and sampled exactly with BP;
in particular, a max-order automaton can forbid overlong copied patterns,
including patterns longer than the formal Markov order. This paper shows that
the same automaton/BP interface can be used more softly, adaptively, and
symmetrically.
A weighted recurrence automaton computes a non-negative recurrence activation
\(R\), for any specified family of target patterns, and BP samples exactly from
\(P_\beta(x)\propto P_0(x)\exp(\beta R(x))\). Negative coupling
\((\beta<0)\) makes the target patterns progressively costly during sampling;
positive coupling \((\beta>0)\) rewards the same patterns and turns them into
controlled attractors. The target patterns need not be the most common patterns
in the generated history: they may be mined from the history, supplied by a
score, derived from a style vocabulary, or designed as an experimental probe.
The core experiments instantiate the general mechanism homeostatically, by
choosing patterns that become overactive in the sampling history. Monophonic
symbolic music is used as a demanding testbed because recurrence is audible,
measurable, and structurally meaningful. Controlled long-horizon comparisons on
six duration-bearing monophonic sources, including Bach and Telemann material,
show that this negative branch reduces generated 8-gram self-reuse, increases
the effective number of generated 8-grams, and increases coverage of
training-supported 4-gram contexts, while preserving substantial lower-order
support. A pitch-sequence replication on five Weimar Jazz Database jazz solos
gives the same anti-reuse signature outside Baroque material. The mechanism
provides signed control of the generator's recurrence landscape: negative
weights provide homeostatic anti-collapse control, while positive weights
provide an exact way to probe attractor basins, phase transitions, and
hysteresis in the underlying variable-order model.
\end{abstract}

\section{Introduction}

Symbolic sequence generation often requires more than locally plausible
next-symbol decisions. A continuation may fit the learned short-range
statistics while entering a recurring corridor: an overused suffix, a repeated
motif, a locally periodic passage, or a copied fragment longer than the model's
formal order. This problem is especially visible in high-order and
variable-order Markov models, where longer contexts improve local specificity
but can also make long-horizon self-reuse more likely. Reducing memory or
adding noise may break these repetitions, but it also weakens the structure
that made the generator useful.

Regular constraints provide a precise interface for controlling such behavior.
In BP-Regular Markov generation, finite automata encode admissible continuations
and belief propagation samples exactly from the Markov distribution conditioned
on automaton acceptance \cite{papadopoulos2015bp}. This framework has already
been used for hard endpoint and avoidance constraints, including max-order
automata that forbid copied chunks exceeding a chosen length
\cite{papadopoulos2014maxorder}. A recognizer can also define a graded
preference: a pattern can be discouraged, rewarded, or swept continuously from
neutral to repulsive to attractive.

This paper introduces signed pattern control. A regular recognizer computes an
activation \(R(x)\) for a candidate continuation \(x\), and the sampler draws
from a reweighted distribution
\[
P_\beta(x) \propto P_0(x)\exp(\beta R(x)).
\]
Negative coupling \((\beta<0)\) penalizes the recognized pattern family,
positive coupling \((\beta>0)\) rewards it, and \(\beta=0\) recovers the
unmodified generator. The target family may be a motif, cadence, rhythm,
contour, syntactic fragment, hand-authored taboo, shuffled control, or a family
mined online from the generator itself.

This separation between pattern selection and signed sampling is central. The
experiments instantiate one policy homeostatically, by choosing patterns that
become overactive during generation, and show that negative weights reduce
long-horizon tunneling without changing the source model. A complementary
fixed-motif probe shows that the same recognizer can be held neutral, made
repulsive, or made attractive. Positive weights then become an experimental
instrument for probing attractor basins.

The idea that absence can itself be musically informative has a precedent in
Conklin's work on antipatterns: patterns that are rare, under-represented, or
absent in a piece, corpus, or genre, and can serve as descriptive or
discriminative musical features \cite{conklin2013antipattern}. Conklin and
Weisser further apply antipattern discovery to Ethiopian bagana songs, where
rare motifs can reveal structural properties of a musical style
\cite{conklinweisser2016bagana}. The present use is complementary: a specified
regular recognizer becomes a signed generative energy and is sampled exactly
within the BP interface.

Monophonic symbolic music is used here as a demanding testbed because
recurrence is audible, measurable, and structurally meaningful. The construction
applies beyond music: any discrete sequence domain with regular pattern
families and a tractable constrained sampler can use the same signed control
principle. In creative domains, signed pattern control steers and probes a
generator's recurrence landscape; proposal and selection remain separate
modelling components. This connects to computational-creativity evaluation
criteria, where novelty is balanced against typicality or value
\cite{ritchie2007}, and to learning-progress accounts that distinguish
arbitrary surprise from learnable regularity \cite{schmidhuber2010}.

\section{BP-Regular Variable-Order Generation}

Belief propagation (BP) is the sum-product message-passing algorithm used to
compute marginal weights in a factor graph. When the factor graph is a chain or
a tree, BP is exact: messages propagated forward and backward give the exact
normalizing mass and the exact conditional distributions needed for sequential
sampling. Papadopoulos, Pachet, Roy, and Sakellariou used this observation to
sample Markov sequences under Regular constraints by building a linear factor
graph whose binary factors combine a Markov transition with the transition of a
finite automaton \cite{papadopoulos2015bp}. BP-Regular sampling draws exactly
from the Markov distribution conditioned on acceptance by one or more automata.

The present generator keeps this BP-Regular interface but uses a variable-order
Markov model. The exact variable-order extension is the sparse context-state
construction of Pachet \cite{pachet2026voregularbp}: BP runs on the observed
context states retained by the variable-order/backoff model and then crosses
them with the regular constraint automaton. Given a current history \(h\), the
variable-order model supplies a distribution over the next symbol by backing
off through an order stack:
\[
p(x_t \mid h_t)
  =
  p_{o_t}(x_t \mid x_{t-o_t:t-1}),
\]
where \(o_t\) is the highest admissible context order retained by the model and
its backoff policy. In ordinary sampling, this distribution is used directly.
In BP-regular sampling, one instead samples a short future
\(x_{1:T}\) subject to a regular constraint. A deterministic finite automaton
\(A=(Q,q_0,\delta,F)\) tracks whether the sampled future remains admissible:
\[
q_t = \delta(q_{t-1},x_t),
\qquad
q_T \in F.
\]
The hard constrained distribution is:
\[
P(x_{1:T}\mid h,A)
  =
  \frac{1}{Z(h,A)}
  \left(
    \prod_{t=1}^{T} p(x_t \mid h,x_{1:t-1})
  \right)
  \mathbf{1}[q_T \in F],
\]
where \(Z(h,A)\) is the normalizing success mass. BP computes and samples from
this constrained product on the reachable context-state/automaton product. The
automaton can change while the underlying Markov model stays fixed: hard ending
constraints, avoidance constraints, and recurrence fields all enter through the
same BP-regular interface.

The recurrence automaton used here is also directly descended from the
max-order constraint introduced by Papadopoulos, Roy, and Pachet
\cite{papadopoulos2014maxorder}. That paper observed that generation can copy
chunks longer than the Markov order used for training: higher-order Markov
generators can reproduce corpus fragments much longer than their formal order.
Max-order addresses this by constructing an automaton that recognizes sequences
avoiding forbidden chunks longer than a chosen maximum.
The present paper keeps the same automaton-centered view of overlong recurrence,
and replaces the hard maximum-order ban with a signed soft energy over
currently overactive generated patterns inside the BP-Regular sampler.

\section{From Hard Acceptance to Signed Soft Acceptance}

A hard regular constraint accepts or rejects a continuation. Equivalently, it
assigns a binary weight:
\[
w_{\mathrm{hard}}(x_{1:T})
  =
  \mathbf{1}[q_T \in F].
\]
The recurrence mechanism uses the same structure but replaces binary
acceptance by a non-negative path weight. The automaton accumulates a
non-negative recurrence activation \(R(x_{1:T})\): the amount of target
recurrence completed by the candidate future. A scalar coupling \(\beta\)
then gives the soft weight
\[
w_{\beta}(x_{1:T})
  =
  \exp[\beta R(x_{1:T})].
\]
The sampled distribution becomes:
\[
P_{\beta}(x_{1:T}\mid h,A)
  =
  \frac{1}{Z_{\beta}(h,A)}
  \left(
    \prod_{t=1}^{T} p(x_t \mid h,x_{1:t-1})
  \right)
  \exp[\beta R(x_{1:T})].
\]
A zero coupling recovers ordinary BP/VO sampling. Negative coupling
\((\beta<0)\) is the homeostatic regime: repeated patterns are discouraged,
but BP may still choose them when all other constraints and local probabilities
make them appropriate. Positive coupling \((\beta>0)\) is the attractor
regime: the same detected patterns are rewarded. Hard rejection remains a
separate limiting case obtained by assigning zero weight to disallowed
transitions.

The recurrence field can also be non-adaptive. Formally, \(R\) can be supplied
by any weighted finite-state recognizer: it may recognize a fixed motif, a
hand-authored taboo, a corpus-derived cliche, a rhythmic cell, or a family of
patterns mined online from the current generation. The homeostatic experiments
below use the last option, because their goal is to control self-tunneling.

Signed softening preserves exact BP-Regular sampling. The boolean automaton
factor is replaced by a weighted automaton factor, and the chain factor graph
still contains only local non-negative potentials. Sum-product BP computes the
exact normalizing mass \(Z_{\beta}(h,A)\) and the exact conditional marginals
for the reweighted distribution \(P_{\beta}\). Since \(\exp[\beta R]\) is
always positive, both signs of \(\beta\) define valid probability
distributions.

The effect of the sign is immediate from the odds ratio. For two candidate
futures \(x\) and \(y\),
\[
\log
\frac{P_{\beta}(x\mid h,A)}{P_{\beta}(y\mid h,A)}
=
\log
\frac{P_0(x\mid h,A)}{P_0(y\mid h,A)}
+
\beta\left(R(x)-R(y)\right).
\]
If \(x\) completes more of the detected recurrence field than \(y\), then a
positive \(\beta\) increases its log-odds by exactly
\(\beta(R(x)-R(y))\), while a negative \(\beta\) decreases those log-odds by
the same amount. Penalty and reward are two signs of the same exact BP-Regular
energy.

For a set \(G_t\) of target patterns active at time \(t\), the transition
activation is:
\[
R_t
  =
  \sum_{g \in G_t}
  \kappa_t(g)
  \mathbf{1}[
    \mathrm{suffix}_{|g|}(x_{1:t}) = g
  ],
\qquad
R(x_{1:T})=\sum_{t=1}^{T} R_t.
\]
Thus the signed recurrence field is just a weighted finite-state recognizer of
target suffixes. In the adaptive homeostatic instantiation, \(G_t\) is chosen
from recently overactive generated suffixes; in an attractor probe, \(G_t\) may
instead be fixed or scenarized externally.

\subsection{Reproducibility}

The implementation, MIDI sources, experiment scripts, and generated paper
assets are available in the Transformator repository:
\url{https://github.com/fpachet/transformator}. The homeostatic replication
tables are produced by
\path{scripts/run_penalty_closing_experiment.py}; the source MIDI files used
by the duration-bearing panels are in \path{data/source_midis/}, and the score
excerpts used in the tunnel examples and fixed-motif probe are included under
\path{docs/assets/}. The repository records the random seeds, query positions,
target lengths, BP order, soft-pattern parameters, and trace diagnostics used
to compute the reported self-reuse, coverage, loss, and computational-overhead
measures.

\section{Adaptive Homeostatic Pattern Injection}

The experiments in this paper instantiate the general signed field as an
adaptive recurrence memory over generated material. At each generated symbol,
the memory is updated through one or more projections. In the music experiments
below, events are projected to pitch class:
\[
\pi(x)=\mathrm{pitch}(x)\bmod 12.
\]
For each order \(k\) in a finite set \(K=\{2,3,4,6,8\}\), the mechanism counts
projected \(k\)-grams in a recent window and in a lifetime stream:
\[
n^{\mathrm{recent}}_t(g),
\qquad
n^{\mathrm{life}}_t(g).
\]
Only grams whose counts exceed minimum recurrence thresholds are turned into
active recurrence patterns. For a pattern \(g\) of order \(k\), the implemented
activation strength has a recent part and a lifetime part:
\[
\rho_t(g)
  =
  \mathrm{recent}(g)
  +
  \mathrm{lifetime}(g).
\]
The recent contribution is normalized within an order so that the most
overactive recent pattern at that order receives the full order-weighted
pressure:
\[
\mathrm{recent}_t(g)
  =
  \lambda_r
  \frac{k}{k_{\max}}
  \frac{
    n^{\mathrm{recent}}_t(g)-m_r+1
  }{
    \max_{g': |g'|=k}
    \left(n^{\mathrm{recent}}_t(g')-m_r+1\right)
  }.
\]
The lifetime contribution grows more slowly, using a power law:
\[
\mathrm{lifetime}_t(g)
  =
  \lambda_{\ell}
  \frac{k}{k_{\max}}
  \left(n^{\mathrm{life}}_t(g)-m_{\ell}+1\right)^{\alpha}.
\]
The total activation is capped, and only the strongest patterns are kept when
there are too many candidates. In the experiments below,
\(\lambda_r=1.5\), \(\lambda_\ell=0.25\), \(m_r=m_\ell=2\),
\(\alpha=0.5\), \(k_{\max}=8\), and at most 96 patterns are injected.

Operationally, every BP decision follows the same cycle:
\begin{enumerate}
  \item read the current generated history;
  \item detect repeated projected grams at several orders;
  \item build a weighted max-order recurrence automaton from these grams;
  \item inject that automaton into the BP graph for the next sampled horizon;
  \item sample the next symbol and update the memory.
\end{enumerate}

The automaton is ``max-order'' in the practical sense that it tracks several
orders at once and lets the highest-order active repetitions carry the strongest
pressure through the factor \(k/k_{\max}\). The result is local and causal: the
generator is pushed away from or toward patterns that have actually become
overactive in its own generated past, depending only on the sign of \(\beta\).

At this level of abstraction, the negative branch is close in spirit to the
``true'' self-avoiding walk, where the walk is biased away from places it has
already visited \cite{amit1983}. Here the repulsion is symbolic and soft. The
positive branch is the matching self-attracting walk: it intentionally makes
revisited symbolic locations more probable.

The mechanism changes the BP energy landscape. Specified high-order corridors
become less attractive or more attractive under exact control; symbol proposal
and transformation remain separate modelling choices.

\section{Homeostatic Experiment}

The empirical tables below evaluate the negative branch of the signed field.
They compare two samplers on the same source, query, Markov model, and random
seed:
\[
\begin{array}{ll}
\text{baseline} & \text{BP/VO sampler, } \beta=0,\\
\text{penalty}  & \text{same sampler plus soft max-order recurrence field, } \beta<0.
\end{array}
\]
The experiment tests whether the negative branch reduces long-horizon
recurrence collapse while preserving lower-order support. The positive branch
is evaluated as an attractor probe after the homeostatic results.

\subsection{Source Panels}

All source files are included with the repository. The main duration-bearing
panel uses MIDI notes as symbols, including pitch and duration. It contains
the two original Bach sources plus four additional monophonic solo-instrument
sources:

\begin{center}
\small
\begin{tabular}{lllrl}
\toprule
panel & source & file & notes & query\\
\midrule
core & Bach Prelude & \path{bach_prelude_c_flat_16ths.mid} & 549 & 448\\
core & Bach Partita & \path{bach_partita_violin.mid} & 1910 & 448\\
clean & Bach BWV 1013, mvt. 1 & \path{bach_bwv1013_mvt1.mid} & 669 & 128\\
clean & Telemann Fantasia 7 & \path{telemann_fantasia7.mid} & 1366 & 128\\
clean & Telemann Fantasia 10 & \path{telemann_fantasia10.mid} & 1108 & 128\\
clean & Telemann Fantasia 12 & \path{telemann_fantasia12.mid} & 1281 & 128\\
\bottomrule
\end{tabular}
\end{center}

A second Bach BWV 1013 movement is included as a diagnostic MIDI file. It is
strictly monophonic, but it did not sustain full-length BP/VO generations from
the current query condition, so it is excluded from the aggregate tables.

As a non-Baroque control, we add five pitch-only solos from the Weimar Jazz
Database (WJazzD) \cite{pfleiderer2017}, following the maximum-entropy
style-modelling abstraction of retaining pitch and replacing duration by a
fixed value \cite{sakellariou2017}. The five solos are Cannonball Adderley,
\emph{So What};
Charlie Parker, \emph{Donna Lee}; David Liebman, \emph{Softly as in a Morning
Sunrise}; John Coltrane, \emph{Giant Steps}; and Miles Davis, \emph{Airegin}.
All five WJazzD MIDI files are strictly monophonic under the scanner used here.
Because this condition is pitch-only and uses BP order 2, its rows are included
for comparison but not pooled with the duration-bearing Bach and Telemann rows.

\subsection{Protocol}

For the duration-bearing panel, the main replication generates 4096 and 8192
events with seeds 17, 23, and 31. A longer single-seed stress check generates
16384 events with seed 17. The parameter sweep uses 4096-event generations with
seeds 17 and 23. The jazz pitch-sequence panel generates 4096 events with seed
17 for each solo. In all cases, the reported measures are computed on the
generated continuation, excluding the query prefix.

Unless otherwise stated, duration-bearing runs use BP order 4 and a one-symbol
BP horizon at each emitted event; the pitch-only WJazzD runs use BP order 2 and
the same one-symbol horizon. The homeostatic penalty uses the parameters given
above: orders \(\{2,3,4,6,8\}\), recent window 128, at most 96 active patterns,
cost cap 8.0, \(\lambda_r=1.5\), and \(\lambda_\ell=0.25\). In implementation,
the coupling magnitude is folded into the non-negative transition cost
\(c_t\): the penalty branch multiplies a candidate transition by
\(\exp(-c_t)\), while the reward branch multiplies it by \(\exp(c_t)\).

\subsection{Measures}

The measures test anti-collapse and exploration:

\begin{itemize}
  \item \textbf{Generated self-reuse:} online reuse rate of generated
  pitch-class 8-grams after first occurrence.
  \item \textbf{Effective 8-gram count:} \(2^H\), where \(H\) is the entropy of
  the generated pitch-class 8-gram distribution.
  \item \textbf{Training-context coverage:} fraction of training pitch-class
  4-grams or 8-grams visited by the generated sequence.
  \item \textbf{Longest repeated suffix:} longest generated pitch-class suffix,
  up to length 32, that had already occurred.
  \item \textbf{Lower-order support:} fraction of generated 2- and 3-grams that
  occur in the source, measured in the pitch-class projection.
  \item \textbf{Static model loss:} mean loss under the source-trained Markov
  model on the full event alphabet used by that panel.
\end{itemize}

\section{Results}

\subsection{Cross-Domain Replication}

Table~\ref{tab:cross-domain-replication} reports the common anti-collapse
metrics across all replication panels. The Bach core rows aggregate the Prelude
and Partita runs. The added monophonic rows aggregate the Bach BWV 1013 and
Telemann runs that reached the requested target length. The WJazzD rows are
shown in the same table as a pitch-only condition with BP order 2 and one seed
per solo.

\begin{table}[t]
\centering
\small
\setlength{\tabcolsep}{3pt}
\begin{tabular}{lllrlrrrrr}
\toprule
panel & repr. & length & \(n\) & condition & self8 & eff8 & cov4 & lower & suffix\\
\midrule
Bach core & dur. & 4096 & 6 & baseline & 0.140 & 3269 & 0.869 & 0.956 & 16.0\\
Bach core & dur. & 4096 & 6 & penalty  & 0.064 & 3728 & 0.909 & 0.937 & 12.5\\
Bach core & dur. & 8192 & 6 & baseline & 0.205 & 5762 & 0.940 & 0.956 & 17.0\\
Bach core & dur. & 8192 & 6 & penalty  & 0.109 & 6963 & 0.963 & 0.934 & 14.5\\
Bach core & dur. & 16384 & 2 & baseline & 0.295 & 9502 & 0.966 & 0.956 & 18.5\\
Bach core & dur. & 16384 & 2 & penalty  & 0.169 & 12617 & 0.990 & 0.929 & 15.5\\
\midrule
Added mono & dur. & 4096 & 12 & baseline & 0.067 & 3630 & 0.855 & 0.950 & 19.5\\
Added mono & dur. & 4096 & 12 & penalty  & 0.035 & 3857 & 0.882 & 0.935 & 15.2\\
Added mono & dur. & 8192 & 12 & baseline & 0.105 & 6720 & 0.945 & 0.950 & 22.2\\
Added mono & dur. & 8192 & 12 & penalty  & 0.054 & 7496 & 0.957 & 0.931 & 16.8\\
Added mono & dur. & 16384 & 4 & baseline & 0.151 & 12231 & 0.984 & 0.949 & 24.0\\
Added mono & dur. & 16384 & 4 & penalty  & 0.083 & 14279 & 0.992 & 0.927 & 17.8\\
\midrule
WJazzD & pitch & 4096 & 5 & baseline & 0.081 & 3606 & 0.887 & 0.963 & 14.8\\
WJazzD & pitch & 4096 & 5 & penalty  & 0.040 & 3862 & 0.917 & 0.950 & 14.0\\
\bottomrule
\end{tabular}
\caption{Cross-domain recurrence-control replication. ``dur.'' rows
use duration-bearing musical events; ``pitch'' rows use the pitch-only WJazzD
abstraction. \(n\) is the number of generated continuations averaged in the
row.}
\label{tab:cross-domain-replication}
\end{table}

The same signature appears in every panel. The penalty reduces generated
pitch-class 8-gram self-reuse, increases the effective number of generated
8-grams, and usually increases training 4-gram coverage. The single-seed
16384-event stress checks preserve the direction of the effect in both
duration-bearing panels. The WJazzD pitch-only rows show the same pattern on a
non-Baroque corpus: self8 drops from 0.081 to 0.040, eff8 rises from 3606 to
3862, and cov4 rises from 0.887 to 0.917.

The cost is also consistent across panels. Lower-order support drops by roughly
one to three percentage points, and the static source-model loss increases in
the measured diagnostics. Source-level stress diagnostics point in the same
direction: in Telemann Fantasia 12 at 16384 events, self8 falls from 0.224 to
0.090, effective 8-gram count rises from 9677 to 14100, and the maximum count
of a generated 8-gram falls from 182 to 34. The mechanism acts as a homeostatic
pressure: less high-order self-reuse and broader exploration, with a measurable
local-likelihood cost.

\subsection{Computational Overhead}

The recurrence field does not add states to the source variable-order model.
Its structural cost is local to each BP call: when the recognizer is active,
the BP problem is crossed with reachable regular states for the current
horizon. Table~\ref{tab:computational-overhead} reports a representative
Prelude seed-17 probe. The source context graph has 1261 context states and
2129 context edges at BP order 4. The ``extra'' columns report the mean
time-indexed regular-product nodes and transitions added per emitted event by
the recognizer. They are zero for the baseline because no regular product graph
is constructed.

\begin{table}[t]
\centering
\small
\begin{tabular}{rlrrrr}
\toprule
length & condition & ms/event & active patterns & extra nodes & extra edges\\
\midrule
4096  & baseline & 0.67 & 0.0  & 0.0  & 0.0\\
4096  & penalty  & 2.75 & 94.2 & 13.0 & 9.7\\
8192  & baseline & 0.71 & 0.0  & 0.0  & 0.0\\
8192  & penalty  & 5.11 & 95.1 & 12.9 & 9.6\\
16384 & baseline & 0.80 & 0.0  & 0.0  & 0.0\\
16384 & penalty  & 10.42 & 95.5 & 12.7 & 9.5\\
\bottomrule
\end{tabular}
\caption{Representative computational overhead for the Prelude seed-17
homeostatic run. Active patterns are the mean number of injected recurrence
recognizers per decision, including early inactive steps. Extra nodes and edges
are the time-indexed reachable regular-product BP graph added by the weighted
recognizer, not new states in the source Markov model.}
\label{tab:computational-overhead}
\end{table}

In this probe, the added graph is small: roughly 13 extra time-indexed nodes
and 10 extra transitions per decision while the recognizer tracks about 95
active patterns. The runtime increase mostly reflects rebuilding the adaptive
recognizer and recomputing the crossed BP problem at each generated event.

\subsection{Strength and Order Sweep}

The parameter sweep uses 4096-event generations with seeds 17 and 23. The
default condition uses orders \(\{2,3,4,6,8\}\), recent strength
\(\lambda_r=1.5\), and lifetime strength \(\lambda_\ell=0.25\).

\begin{center}
\small
\begin{tabular}{llrrrrr}
\toprule
piece & variant & self8 & eff8 & lower & suffix & max8\\
\midrule
Prelude & weak & 0.105 & 3498 & 0.941 & 14.0 & 6.0\\
Prelude & default & 0.105 & 3496 & 0.940 & 14.5 & 5.5\\
Prelude & strong & 0.098 & 3540 & 0.936 & 12.0 & 5.0\\
Prelude & no lifetime & 0.137 & 3308 & 0.951 & 16.0 & 6.0\\
Prelude & high order only & 0.135 & 3328 & 0.952 & 14.0 & 5.5\\
Partita & weak & 0.028 & 3929 & 0.937 & 12.5 & 3.5\\
Partita & default & 0.024 & 3951 & 0.935 & 11.0 & 3.5\\
Partita & strong & 0.024 & 3950 & 0.932 & 13.0 & 4.0\\
Partita & no lifetime & 0.053 & 3769 & 0.947 & 13.5 & 7.0\\
Partita & high order only & 0.043 & 3837 & 0.951 & 12.0 & 5.0\\
\bottomrule
\end{tabular}
\end{center}

The sweep shows no fragile tuned optimum. Weak, default, and strong multi-order
penalties all improve the anti-tunnel profile. Removing the lifetime term or
penalizing only high orders weakens the effect, especially on the Partita. The
mechanism acts as a multi-scale homeostatic memory: recent pressure reacts
quickly, while lifetime pressure prevents the same corridors from gradually
re-entering the generated stream.

\section{Concrete Tunnel Examples}

The aggregate measures are useful, but the mechanism is easier to understand
by looking at repeated passages. The following examples use 4096-note runs.
They compare matched source, prompt, seed set, and sampler; after the first
stochastic divergence, the local positions differ, but the recurrence profiles
remain comparable. The score excerpts were exported as MEI and engraved with
Verovio.

Before inspecting individual Prelude excerpts, we also checked the negative
case in the matched penalty runs. For every 4096-note Prelude run, we scanned
exact repeated pitch and pitch-class $n$-grams up to order 64. The penalty
condition still contains ordinary recurrence while avoiding the baseline-scale
tunnels illustrated below: no repeated pitch-class passage reaches length 16,
and the strongest repeated 8-gram corridors occur at most 6 times.

\begin{center}
\small
\begin{tabular}{lrrr}
\toprule
Prelude run & longest pitch & longest pc & max pc8 count\\
\midrule
baseline 17 & 20 & 20 & 11\\
baseline 23 & 13 & 14 & 7\\
baseline 31 & 18 & 19 & 11\\
penalty 17 & 15 & 15 & 6\\
penalty 23 & 14 & 14 & 5\\
penalty 31 & 13 & 13 & 6\\
\bottomrule
\end{tabular}
\end{center}

The penalty preserves ordinary self-similarity while removing the salient
baseline behavior shown by the examples: deep literal returns and highly
dominant high-order corridors in the long continuation.

\subsection{Seed 17: repeated 8-gram corridor}

In the baseline run, the pitch-class corridor
\[
\texttt{D G B D G B D G}
\]
appears 11 times, at positions
\[
316,642,843,1242,1457,1677,2167,2603,2697,3436,3865.
\]
Two concrete pitch realizations are:
\[
\texttt{D3 G3 B3 D4 G3 B3 D4 G2}
\]
and
\[
\texttt{D4 G3 B3 D4 G4 B4 D4 G4}.
\]
The exact registers differ, but the pitch-class corridor is the same. In the
matched penalty run, the most repeated 8-gram appears only 6 times. The change
globally redistributes repeated high-order pressure across candidate futures.

\begin{figure}[ht]
\centering
\includegraphics[scale=0.44]{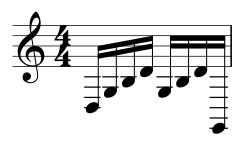}
\hspace{0.15\linewidth}
\includegraphics[scale=0.44]{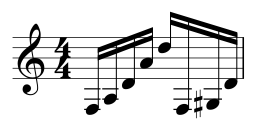}
\caption{Left: one occurrence of the baseline seed-17 8-note corridor that
appears 11 times as a pitch-class pattern. Right: the strongest 8-note
corridor in the matched penalty run, whose maximum count is 6.}
\end{figure}

\subsection{Seed 17: a 20-note literal pitch-class tunnel}

The same baseline run repeats the following 20-note pitch-class passage at
positions 1115 and 2479:
\[
\texttt{C F G Eb A C Eb Ab F A C F A C F\# G E G C E}.
\]
The absolute pitch realization is identical in both occurrences:
\[
\begin{array}{l}
\texttt{C4 F4 G2 Eb3 A3 C4 Eb4 Ab2 F3 A3}\\
\texttt{C4 F4 A3 C4 F\#4 G2 E3 G3 C4 E4}
\end{array}
\]
In the matched penalty run, the baseline length 20 drops to 15:
\[
\texttt{A D A D F Ab D G B G Bb E C C F}.
\]

\begin{figure}[ht]
\centering
\includegraphics[scale=0.44]{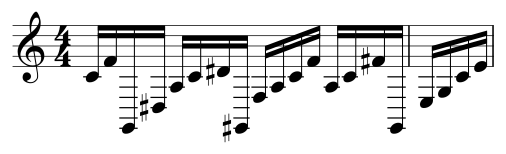}
\hfill
\includegraphics[scale=0.44]{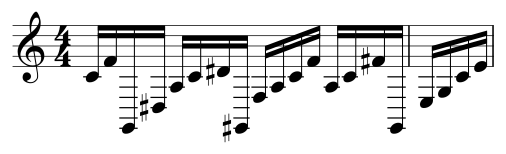}
\caption{The same 20-note absolute-pitch passage appears twice in the baseline
seed-17 run, at positions 1115 and 2479. This is a concrete tunnel visible in
the score.}
\end{figure}

\begin{figure}[ht]
\centering
\includegraphics[scale=0.44]{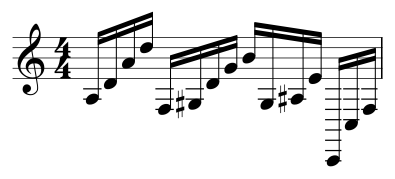}
\caption{The longest repeated passage in the matched seed-17 penalty run has
length 15 pitch classes. The mechanism reduces the dominance and depth of
repeated high-order corridors.}
\end{figure}

\subsection{Seed 31: another high-order corridor}

In baseline seed 31, two overlapping 8-gram corridors each appear 11 times:
\[
\texttt{G B F G D G B D}
\qquad
\text{and}
\qquad
\texttt{B F G D G B D G}.
\]
The first occurs at positions
\[
13,80,381,532,1686,2285,2396,2809,3198,3472,3943.
\]
The baseline also repeats a 19-note pitch-class passage. In the matched penalty
run, the strongest 8-gram count is 6 and the longest repeated passage drops to
13. The mechanism prevents a small number of high-order corridors from
dominating the long continuation.

\subsection{Partita seed 17: transfer to another source}

The Partita source is less prone to long literal recurrence than the Prelude,
but the same qualitative pattern appears. In baseline seed 17, the strongest
pitch-class 8-gram appears 11 times. In the matched penalty run, the strongest
8-gram appears only 3 times.

\begin{figure}[ht]
\centering
\includegraphics[scale=0.44]{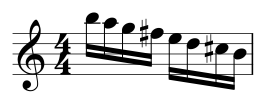}
\hspace{0.15\linewidth}
\includegraphics[scale=0.44]{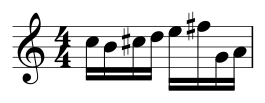}
\caption{Partita seed 17. Left: one occurrence of the strongest baseline
8-gram corridor, which appears 11 times as a pitch-class pattern. Right: the
strongest 8-gram corridor in the matched penalty run, whose maximum count is
3.}
\end{figure}

The baseline run also repeats a 14-note absolute-pitch passage at positions
339 and 652:
\[
\texttt{B4 A4 C\#5 B4 A4 G4 F\#4 E4 D4 C\#4 B3 D4 F\#4 B4}.
\]
In the matched penalty run, the longest repeated pitch passage has length 11.

\begin{figure}[ht]
\centering
\includegraphics[scale=0.44]{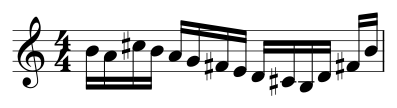}
\hfill
\includegraphics[scale=0.44]{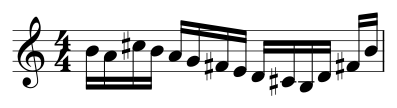}
\caption{The same 14-note absolute-pitch passage appears twice in the Partita
baseline seed-17 run. The matched penalty run reduces the longest repeated
pitch passage to 11 notes.}
\end{figure}

\section{Interpretation}

The experiments support the negative branch. Soft recurrence energies add an
adaptive signed memory to the BP-Regular sampler. With \(\beta<0\), material
that has become overactive receives a higher relative cost inside the exact
weighted BP distribution.

The mechanism is:
\[
\text{signed recurrence field} =
\text{adaptive memory over generated recurrence}.
\]
The negative branch is:
\[
\text{negative recurrence field} =
\text{homeostatic anti-collapse and exploration regulator}.
\]
The positive branch is:
\[
\text{positive recurrence field} =
\text{controlled attractor probe}.
\]

Negative weights make overactive high-order corridors less attractive, so the
generator keeps moving through training-supported contexts instead of
collapsing into a few recurring paths. Positive weights deliberately create
such corridors, so their attraction basins can be measured. The mechanism adds
signed control over the recurrence landscape of the variable-order BP-Regular
generator. In the experiments, adaptive recurrence costs reduce repeated
high-order contexts while preserving much of the lower-order style support.

\section{Attractor Probes and Their Evaluation}

The positive branch uses a different evaluation criterion from the homeostatic
branch. A successful penalty run is less tunnel-like. A successful reward run
can become more tunnel-like; the question is whether this happens in a
controlled, interpretable way. Positive weights provide an experimental
stimulus for probing the VO model.

The general scenarization is a time-varying signed field:
\[
P(x_{1:T}\mid h)
\propto
P_0(x_{1:T}\mid h)
\exp\left(
  \sum_{t=1}^{T}\sum_i \beta_i(t) R_{i,t}(x_{1:t})
\right),
\]
where each \(R_i\) may recognize a different projected pattern family:
literal pitch-class recurrence, interval recurrence, rhythmic recurrence,
cadential fragments, externally specified motifs, or shuffled control
patterns. These recognizers need not be frequent in the generated past; they
can be chosen before generation, updated from the current history, or changed
by an experimental schedule. The schedule \(\beta_i(t)\) is the scenario.
Negative values create repulsion, positive values create attraction, and sign
changes create recovery tests.

\subsection{Signed Fixed-Motif Probe}

A minimal concrete probe uses one externally selected target pattern and three
signs. Choose a plausible four-note pitch-class pattern from the training
source, then generate continuations from the same source, query, and seeds
under:
\[
\beta_{\mathrm{motif}}=0,\qquad
\beta_{\mathrm{motif}}>0,\qquad
\beta_{\mathrm{motif}}<0.
\]
The three conditions can be read as \emph{neutral} (nothing),
\emph{rewarded} (pattern), and \emph{penalized} (antipattern). This isolates
the sign of the same recognizer: the neutral condition measures the motif's
ordinary probability under the VO model, the positive condition asks whether it
can be made attractive, and the negative condition asks whether it can be made
avoidable.

As a small pilot, we used the Bach BWV 1013 first movement and selected the
pitch-class motif
\[
\texttt{E A B C},
\]
which occurs six times in the training sequence and is ranked third among
pitch-class 4-grams. With 128-event continuations, three seeds, BP order 4,
horizon 8, and signed cost magnitude \(3.0\), the same recognizer gives:

\begin{center}
\small
\begin{tabular}{lrrrrrr}
\toprule
condition & motif count & motif rate & lift & self4 & eff4 & lower\\
\midrule
neutral   & 0.7 & 0.005 & 1.00 & 0.117 & 105.7 & 0.937\\
penalized & 0.0 & 0.000 & 0.00 & 0.120 & 104.9 & 0.952\\
rewarded  & 3.3 & 0.026 & 5.00 & 0.184 & 94.7 & 0.952\\
\bottomrule
\end{tabular}
\end{center}

These fixed-sign controls answer the causal question: what changes when the
same recognizer is held neutral, made repulsive, or made attractive from the
beginning of generation? A complementary display makes the scenario temporal:
one continuous generation is divided into consecutive phases, and only the sign
of the same recognizer is changed at the phase boundaries. For visual clarity,
the phase-scheduled example below uses a stronger magnitude:
\[
\beta(t): 0 \;\rightarrow\; -8.0 \;\rightarrow\; +8.0.
\]

\begin{figure}[ht]
\centering
\includegraphics[width=0.22\linewidth]{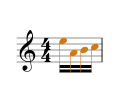}

\vspace{0.5em}
\includegraphics[width=\linewidth]{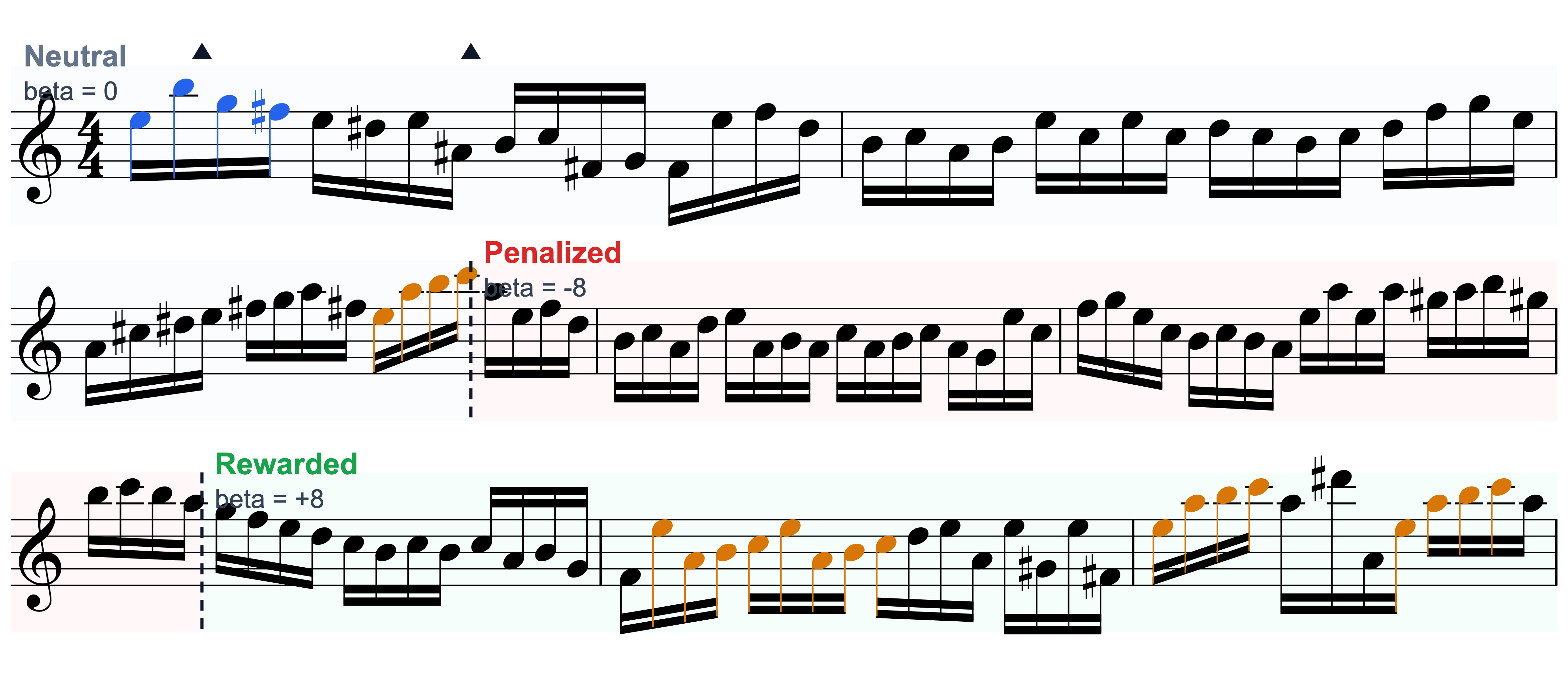}
\caption{Contiguous fixed-motif phase probe, seed 17. Top: the four-note
pitch-class recognizer \texttt{E A B C}, selected from the training sequence (six
source occurrences, rank 3 among pitch-class 4-grams). Bottom: one continuous
query-plus-continuation score, wrapped onto three staff systems. The labels above
the staff mark consecutive 40-note phases with \(\beta=0\), \(\beta=-8\), and
\(\beta=+8\); the vertical dashed lines are the parameter-change instants, not
restarts. Blue notes
indicate the query prefix; orange notes indicate completions of the selected
motif. In this run the motif appears once in the neutral phase, is absent in
the penalized phase, and appears four times in the rewarded phase.}
\end{figure}

The probe evaluates controllability of the recognizer. The symmetry is the
point: the same four-symbol regular recognizer is ordinary, repulsive, or
attractive depending only on the sign. The positive condition increases target
completions while also increasing 4-gram self-reuse and reducing effective
4-gram diversity; the negative condition suppresses the target completions
without collapsing the lower-order support. The contiguous phase display shows
the same idea in a scenario: the generator is first left alone, then
discouraged from completing the target, then pulled toward it.

\subsection{Phase-Transition Sweep}

The simplest attractor probe sweeps a single coupling from negative to positive:
\[
\beta \in \{-4,-2,-1,0,1,2,4,8\}.
\]
For each fixed source, query, and seed, the same recurrence recognizer is used
throughout. The evaluation is the response curve, not one run in isolation.
Useful measures are:
\begin{itemize}
  \item \textbf{Target activation per note:} mean \(R_t\), or equivalently the
  mean matched recurrence activation reported by the weighted automaton.
  \item \textbf{Self-reuse response:} generated pitch-class 4- and 8-gram
  self-reuse as a function of \(\beta\).
  \item \textbf{Diversity collapse:} effective 8-gram count and longest
  repeated suffix.
  \item \textbf{Style support cost:} lower-order support and static model loss.
  \item \textbf{Critical coupling:} the smallest positive \(\beta\) at which a
  sharp increase in target activation or a sharp drop in diversity occurs.
\end{itemize}
The response curve locates where the VO/BP model changes regime from ordinary
continuation to biased style to obsessive attractor.

\subsection{Addiction and Withdrawal}

A second probe turns the reward on and then off:
\[
\beta(t)=
\begin{cases}
\beta_+ & 1\le t\le T_1,\\
\beta_- & T_1 < t\le T.
\end{cases}
\]
The first phase deliberately pulls the generated context into a recurrent
basin. The second phase asks whether homeostasis can pull it back out. The
main evaluation measure is \textbf{recovery time}: the number of events after
the sign flip needed for self-reuse, target activation, or entropy-window
statistics to return to the baseline or negative-branch band. If two runs have
the same final negative coupling but different positive prehistories, their
difference measures hysteresis: the generated context itself has become part
of the attractor.

\subsection{Selective and Contradictory Attractors}

The recurrence field can be split into several recognizers. For example,
\[
\beta_{\mathrm{cadence}}>0,\qquad
\beta_{\mathrm{literal\ copy}}<0
\]
rewards cadential or cliche-like fragments while still penalizing literal
self-copy. This tests whether the system can become stylistically attracted
without collapsing into exact reuse. Conversely, one may reward a high-order
pattern while penalizing some lower-order components needed to reach it. Such
contradictory fields evaluate how BP distributes compromise across the future
horizon.

These scenarios use both \textbf{specificity} and \textbf{collateral damage}
measures. Specificity is the fraction of increased
recurrence activation belonging to the intended recognizer. Collateral damage
is the increase in unrelated self-reuse, periodic runs, or static model loss.
A useful positive-control experiment rewards the real target recognizer; a
useful negative control rewards a shuffled or randomly selected recognizer with
the same activation budget. A successful probe produces stronger target
activation for the real recognizer than for the shuffled control at comparable
style-support cost.

\subsection{Temporal Scenarization}

Finally, the signs can be scenarized over time. In a musical testbed, this may
mean rewarding recurrence to create obsession, then penalizing recurrence to
force release; rewarding rhythmic recurrence while penalizing pitch-class
recurrence; or alternating attraction and repulsion every phrase-length block.
In other symbolic domains, the same idea becomes a scheduled policy over
recognized structures. These controlled interventions can be evaluated by
plotting activation, self-reuse, effective diversity, and style-support curves
over time, aligned to the scheduled sign changes. A convincing scenario shows
that the measured recurrence profile follows the intended sign schedule while
remaining inside a usable lower-order support band.

\section{Replication and Extension}

The replication panel covers six duration-bearing monophonic sources: a Bach
keyboard Prelude, a Bach solo violin Partita, one movement from the Bach solo
flute Partita BWV 1013, and three Telemann solo flute Fantasias. For each
source we generate 4096 and 8192 event continuations with three seeds, and we
add a single-seed 16384 event stress check. A separate pitch-sequence
replication covers five WJazzD jazz solos under a pitch-only abstraction. The
result spans pieces, composers, Baroque textures, and a non-Baroque
pitch-sequence condition. Recurrence pressure can be controlled as generation
length increases.

Together, these replications establish the central claim for the tested
negative branch: the mechanism provides explicit, measurable control of
long-horizon recurrence pressure while preserving the underlying
variable-order Markov/BP generator. The signed formulation extends this claim
from stabilization to probing. Negative weights isolate anti-collapse pressure
before any explicit novelty generator is added; positive weights supply a
controlled way to create and measure attractors before interpreting them in a
domain-specific way.

\section{Limitations}

The experiments are intentionally narrow. They use monophonic symbolic
sequences, not polyphonic scores, audio, or expressive performance data. The
duration-bearing panel covers several Bach and Telemann sources, and the
pitch-only WJazzD panel checks transfer outside Baroque material, but the
results should not be read as a broad survey of musical style modelling. They
show that the signed regular field controls recurrence pressure in this tested
VO/BP setting.

The strongest empirical evidence concerns the negative, homeostatic branch.
The positive branch is presented mainly as a controlled probe: it demonstrates
that the same recognizer can be made attractive and suggests phase-transition,
hysteresis, and scenarized-control evaluations. It is not yet a complete
account of musical value, preference, or compositional quality. Finally, the
current implementation rebuilds an adaptive weighted recognizer at each
decision; the overhead is modest in the one-symbol-horizon probes reported
here, but longer horizons and richer recognizer families will require more
careful engineering.

\section{Conclusion}

This paper reframes recurrence control in variable-order Markov/BP generation
as a signed regular energy. A weighted automaton recognizes a chosen family of
target patterns, and BP samples exactly from the corresponding reweighted
distribution. Changing the sign of the coupling changes the behavior but not
the mechanism: negative weights penalize target activations, positive weights
reward them, and both remain exact BP-Regular sampling problems.

The target patterns need not be the most frequent patterns in the generated
past, or even patterns observed in that past at all. They can be supplied
externally, designed by hand, learned from a corpus, scheduled as probes, or
mined adaptively from overactive generated material. The experiments choose the
adaptive policy and show, on symbolic music continuations, that the negative
branch reduces long-horizon self-reuse while preserving substantial lower-order
support.

The positive branch opens the complementary direction: it makes attraction
programmable. By rewarding selected regular patterns, one can deliberately
create attractors, sweep their strength, measure phase transitions, test
recovery after sign reversal, and compare real targets with shuffled controls.
This makes the signed VO/BP construction a general tool for studying and
steering the recurrence landscape of symbolic sequence generators, with music
serving here as a concrete and audibly inspectable domain.


\begin{thebibliography}{14}

\bibitem{amit1983}
D. J. Amit, G. Parisi, and L. Peliti.
\newblock Asymptotic behavior of the ``true'' self-avoiding walk.
\newblock \emph{Physical Review B}, 27(3):1635--1645, 1983.
\newblock \url{https://doi.org/10.1103/PhysRevB.27.1635}.

\bibitem{papadopoulos2015bp}
A. Papadopoulos, F. Pachet, P. Roy, and J. Sakellariou.
\newblock Exact sampling for regular and Markov constraints with belief
propagation.
\newblock In \emph{Principles and Practice of Constraint Programming}, 2015.

\bibitem{pachet2026voregularbp}
F. Pachet.
\newblock Exact regular-constrained variable-order Markov generation via sparse
context-state belief propagation.
\newblock \emph{arXiv preprint arXiv:2605.07839}, 2026.
\newblock \url{https://doi.org/10.48550/arXiv.2605.07839}.

\bibitem{papadopoulos2014maxorder}
A. Papadopoulos, P. Roy, and F. Pachet.
\newblock Avoiding plagiarism in Markov sequence generation.
\newblock In \emph{Proceedings of the Twenty-Eighth AAAI Conference on
Artificial Intelligence}, 2014.

\bibitem{conklin2013antipattern}
D. Conklin.
\newblock Antipattern discovery in folk tunes.
\newblock \emph{Journal of New Music Research}, 42(2):161--169, 2013.
\newblock \url{https://doi.org/10.1080/09298215.2013.809125}.

\bibitem{conklinweisser2016bagana}
D. Conklin and S. Weisser.
\newblock Pattern and antipattern discovery in Ethiopian bagana songs.
\newblock In D. Meredith, editor, \emph{Computational Music Analysis},
pages 425--443. Springer, 2016.
\newblock \url{https://doi.org/10.1007/978-3-319-25931-4_16}.

\bibitem{ritchie2007}
G. Ritchie.
\newblock Some empirical criteria for attributing creativity to a computer
program.
\newblock \emph{Minds and Machines}, 17:67--99, 2007.
\newblock \url{https://doi.org/10.1007/s11023-007-9066-2}.

\bibitem{sakellariou2017}
J. Sakellariou, F. Tria, V. Loreto, and F. Pachet.
\newblock Maximum entropy models capture melodic styles.
\newblock \emph{Scientific Reports}, 7:9172, 2017.
\newblock \url{https://doi.org/10.1038/s41598-017-08028-4}.

\bibitem{schmidhuber2010}
J. Schmidhuber.
\newblock Formal theory of creativity, fun, and intrinsic motivation
(1990--2010).
\newblock \emph{IEEE Transactions on Autonomous Mental Development},
2(3):230--247, 2010.
\newblock \url{https://doi.org/10.1109/TAMD.2010.2056368}.

\bibitem{pfleiderer2017}
M. Pfleiderer, K. Frieler, J. Abe{\ss}er, W.-G. Zaddach, and B. Burkhart,
editors.
\newblock \emph{Inside the Jazzomat: New Perspectives for Jazz Research}.
\newblock Schott Campus, 2017.
\newblock Weimar Jazz Database: \url{https://jazzomat.hfm-weimar.de/}.

\bibitem{ron1996}
Ron, D., Singer, Y., and Tishby, N.
\newblock The power of amnesia: learning probabilistic automata with variable
memory length.
\newblock \emph{Machine Learning}, 1996.

\bibitem{pearce2004}
Pearce, M. T. and Wiggins, G. A.
\newblock Improved methods for statistical modelling of monophonic music.
\newblock \emph{Journal of New Music Research}, 2004.

\end{thebibliography}
\end{document}